\documentclass{article}
\usepackage{spconf,amsmath,graphicx}

\input{mysymbol.sty}

\usepackage[utf8]{inputenc} 
\usepackage[T1]{fontenc}    
\usepackage{hyperref}       
\usepackage{url}            
\usepackage{booktabs}       
\usepackage{amsfonts}       
\usepackage{nicefrac}       
\usepackage{microtype}      
\usepackage{lipsum}
\usepackage{amsmath}
\usepackage{amssymb}        
\usepackage{amsthm} 
\theoremstyle{plain}

\theoremstyle{definition}

\usepackage{verbatim}
\usepackage{empheq,float}
\usepackage{graphicx,caption,subcaption,balance}
\usepackage{stfloats}
\usepackage[dvipsnames]{xcolor}

\newcommand{\sn}{\smallskip\noindent}

\usepackage{epstopdf}
\usepackage{algorithm}
\usepackage{algpseudocode}
\usepackage{mathrsfs}
\usepackage{psfrag}

\title{Forecasting graph Signals with Recursive MIMO Graph Filters}
\name{Jelmer van der Hoeven, Alberto Natali and Geert Leus\thanks{E-mails:\{Jelmer.van.der.hoeven@pwc.com\} \{a.natali;g.j.t.leus\}@tudelft.nl; }}
\address{Faculty of Electrical Engineering, Mathematics and Computer Science\\ Delft University of Technology, Delft, The Netherlands}

\begin{document}
\ninept
\maketitle
\begin{abstract}
 Forecasting time series on graphs is a fundamental problem in graph signal processing. When each entity of the network carries a vector of values for each time stamp instead of a scalar one, existing approaches resort to the use of product graphs to combine this multidimensional information, at the expense of creating a larger graph. In this paper, we show the limitations of such approaches, and propose extensions to tackle them. Then, we propose a recursive multiple-input multiple-output graph filter which encompasses many already existing models in the literature while being more flexible. Numerical simulations on a real world data set show the effectiveness of the proposed models.

\end{abstract}
\section{Introduction}

Forecasting time series collected by entities of a network is a central problem in graph signal processing (GSP)~\cite{Emerging2012}, finding applications in sensor \cite{sensor_example} and social networks~ \cite{social_example,social_example_2}, to name a few. Accounting for the network structure in the forecasting model serves as an inductive bias to reduce the number of estimation parameters of the model~\cite{GSP_Causal2016,GVARMA2019} compared to classical vector autoregressive (VAR) models~\cite{lütkepohl_2005}.  However, existing models under a GSP lens consider a scalar value at each node for each time instant~\cite{KKFTemperature2017,Adaptive_2018,GVARMA2019}, which is a limitation in networks where a (feature) vector of measurements is available at each node for each time instant, such as in meteorological sensor networks or $3$D point clouds.

 This \textit{multidimensional} case has been recently  addressed in~\cite{GVAR_MULTI2020} with the introduction of a so-called feature graph, capturing the possible relationships among the features of a node of the network. There, the authors make use of product graphs~\cite{BigData2014} to create a new larger graph that combines in a principled way the original and the feature graphs, which is then used to forecast these multidimensional content by taking into account relationships among features of different nodes. However, the knowledge of which product graph to use and how to construct the feature graph has not been properly elaborated. 
 
This work can be considered an extension of~\cite{GVAR_MULTI2020} in that: \textit{i)} we first delineate drawbacks of the proposed model and introduce minor extensions to tackle them;  then, \textit{ii)} we propose a recursive multiple-input multiple-output (MIMO) graph filter which generalizes the already existing filters while being more flexible; finally, \textit{iii)} we put forth  a method which learns the weights of the feature graph during the training process.
We limit our exposition to linear models, although non-linear autoregressive models are a subject worth investigating. Numerical experiments on real-world data show the effectiveness of the proposed approaches.

\section{Background}
\vspace{-3mm}
\label{sec:preliminaries}

Consider an $N$-dimensional time series $\bbx_t \in \reals^N$, where entry $x_t(i)$ represents the value at time $t$ associated to the entry $i$. For instance, $x_t(i)$ might represent the amount of water in the $i$th reservoir of a water network at time $t$. When the relationships between the different time series of $\bbx_t$ can be captured by a network, let the graph $\mathcal{G}:=\left(\mathcal{V},\mathcal{E}, \bbS\right)$ model such network, where $\mathcal{V}=\{v_1,  \ldots,v_N\}$ is the set of $N$ nodes, $\mathcal{E} \subseteq \mathcal{V} \times \mathcal{V}$ is the  set of edges, and   $\bbS \in \reals^{N \times N}$ is the matrix representation of the graph, which captures its sparsity pattern; i.e.,  $S_{ij} \neq 0$ if and only if $(v_i,v_j) \in \ccalE$ or $v_i=v_j$. We refer to matrix $\mathbf{S}$ as the graph shift operator (GSO), examples of which include the adjacency matrix and the Laplacian matrix~\cite{DiscreteGSP2014Freq,Overview2018}. In this context, $\bbx_t$ is called a graph signal.

We can instantly process a graph signal $\bbx_t$ to obtain a new graph signal $\bby_t$ by means of the so-called graph convolution~\cite{Emerging2012}:

\begin{equation}
\bby_t=\sum_{k=0}^{K-1} h_{k} \mathbf{S}^{k} \bbx_t,
\label{eq:filtering}
\end{equation}
where $\bbH(\bbS):=\sum_{k=0}^{K-1} h_{k} \mathbf{S}^{k}$ is the graph filter~\cite{DiscreteGSP2013} of order $K-1$ with scalar coefficients $h_0, \ldots, h_{K-1}$. Because matrix $\mathbf{S}$ is sparse, the computational complexity of the graph filtering operation~\eqref{eq:filtering} is $\mathcal{O}(|\mathcal{E}|K)$. 
To forecast time series residing on the nodes, a graph vector autoregressive (G-VAR) model has been introduced in~\cite{GVARMA2019} as the combination of a finite number of graph convolutions; specifically:
\begin{equation}
\mathbf{x}_{t}=-\sum_{p=1}^{P} \mathbf{H}_{p}(\mathbf{S}) \mathbf{x}_{t-p}=-\sum_{p=1}^{P} \sum_{k=0}^{K-1} h_{k p} \mathbf{S}^{k} \mathbf{x}_{t-p},
\label{eq:graph-VAR}
\end{equation}
where $\bbx_{t-p}$ represents the graph signal at time instant $t-p$ and $h_{kp}$ represents the $k$th scalar filter coefficient of the $p$-th graph filter $\bbH_p(\bbS)$. The computational complexity of~\eqref{eq:graph-VAR} is $\ccalO(PK|\ccalE|)$ and the number of parameters of the filter is $PK$, both independent of the size of the network.

  \sn \textbf{Multidimensional.} 
When each node of the network carries a vector of $F$ values (features)  for each time instant $t$, we refer to such a graph signal as being $F$-dimensional,  and denote it as  $\bbx_t= [\bbx_t^\top(1), \ldots, \bbx_t^\top(F)]^\top\in \mathbb{R}^{NF}$, where each $\bbx_t(f) \in \reals^N$ is the one-dimensional graph signal associated to feature $f$. In other words, $\bbx_t$ is the concatenation of $F$ graph signals, each one representing the graph signal associated to one specific feature.

\sn
To efficiently deal with such multi-dimensional graph signals, the work in~\cite{GVAR_MULTI2020} proposes to model the dependencies among the features with a so-called \textit{feature graph} defined as $\mathcal{G}_\mathcal{F}:=\left(\mathcal{V}_\mathcal{F},\mathcal{E}_\mathcal{F}, \bbS_\ccalF \right)$, where $\mathcal{V_F}=\{f_1,....,f_F\}$ is the  set of $F$ nodes representing the features,  $\mathcal{E_F} \subseteq \mathcal{V_F} \times \mathcal{V_F}$ is the  set of edges defining how the features are connected, and $\bbS_\ccalF$ is the associated GSO. To formally capture the dependencies among features of different nodes,  $\mathcal{G}$ and $\mathcal{G}_\mathcal{F}$ can be combined with the use of product graphs~\cite{Handbook2011} to create a new graph $\mathcal{G}_{\diamond}=\left(\mathcal{V}_\diamond,\mathcal{E}_\diamond, \bbS_\diamond \right)$, with node set $\ccalV_\diamond$ of cardinality $|\mathcal{V}_{\diamond}|=NF$,    edge set $\ccalE_\diamond \subseteq \mathcal{V}_\diamond \times \mathcal{V}_\diamond$ and the $NF \times NF$ graph shift operator $\bbS_\diamond$. Its sparsity pattern depends by the type of product graph adopted. A novel aspect of the work in~\cite{GVAR_MULTI2020} comes from the realization that all common types of product graphs,  such as the Kronecker and the Cartesian, can be parametrized as:
\begin{equation}
\label{eq:para}
\mathbf{S}_{\diamond}=\sum_{i=0}^{1} \sum_{j=0}^{1}  s_{i j}\left(\mathbf{S}_{\mathcal{F}}^{i} \otimes \mathbf{S}^{j}\right),
\end{equation}
where $s_{i j} \in\{0,1\}$, which  creates a level of abstraction on the specific product-graph choice for $\bbS_\diamond$. Given a particular product graph $\bbS_\diamond$, according to \eqref{eq:para}, a product graph filter  $\bbH(\cdot)$ can be defined as:
\begin{equation}
\label{eq:PG-maybe}
    \bbH(\bbS_\diamond) =\sum_{k = 0}^{K-1} {h}_{k} \bbS_\diamond^k.
\end{equation}
This filter is then used to forecast the $F$-dimensional graph process $\bbx_t$ as:
\begin{equation}
\label{eq:gpgp-var}
    \bbx_{t}=-\sum_{p=1}^{P} \sum_{k = 0}^{K-1} h_{kp} \mathbf{S}_\diamond^k \bbx_{t-p},
\end{equation}
which is termed product graph VAR (PG-VAR) filter, yielding a total of $PK$ number of parameters.

\sn
Despite the fact that the PG-VAR takes the information on related features into account, it has some limitations. The first is that it uses a limited amount of parameters independent of the number of features, which might be restrictive. Secondly, even though the edge weights $\ccalE$ and $\ccalE_\ccalF$ may well describe the strength of the relationships in the original graph $\ccalG$ and in the feature graph $\ccalG_\ccalF$, their product-graph combination may not. In order to overcome those limitations, in Section~\ref{sec:New_Models} we propose two new graph-based VAR models with an increased amount of flexibility, and in Section~\ref{sec:param_estimation}  we discuss how to optimally learn the feature graph weights during the training of the forecasting process.


\section{An Overarching Forecasting Model}
\label{sec:New_Models}
In this section, we propose two autoregressive graph-based models, which encompass the already existing ones in the literature, tackle the mentioned limitations and bring increased flexibility. The first is a direct extension of the filters introduced in Section~\ref{sec:preliminaries}; the latter is a generalization of the multiple-input multiple-output GF which eliminates the need for knowledge of the feature graph.

\sn
\textbf{Combined. }
Consider the PG-VAR in~\eqref{eq:gpgp-var} and consider the term associated to the index  $k=0$, i.e.,:
\begin{align}
\mathbf{x}_{t}&=-\sum_{p=1}^{P} h_{0 p}\left(\bbI_F \otimes \bbI_N\right)\bbx_{t-p}.
 \label{eq:Parametric_PG_VAR}
\end{align} 
The sum in~\eqref{eq:Parametric_PG_VAR} simply represents  $NF$ uni-variate auto-regressive filters, where all filters are constrained to have the same set of parameters $h_{0p}$. This might be a restriction since it weighs similarly all the features in every node.

For this reason, we first  propose a minor extension of~\eqref{eq:gpgp-var} by adapting its zero-th order coefficients, with a set of  $F$ feature-dependent parameters $h^{(f)}_{0p}$. That is:
\begin{align}
\label{eq:first}
\mathbf{x}_{t}=-\left(\sum_{p=1}^{P}\left( \operatorname{diag}(\mathbf{h}_{0p}) \otimes \mathbf{I}_{N}\right)+ \sum_{p=1}^{P} \sum_{k=1}^{K-1} h_{kp} \mathbf{S}_\diamond^{k}\right) \mathbf{x}_{t-p},
\end{align}
where $\mathbf{h}_{0p} = [h^{(1)}_{0p},\hdots,h^{(F)}_{0p}]^\top \in \reals^F$, and $h^{(f)}_{0p}$ represents the zero-th order filter coefficient  associated to the $f$th feature of the $p$th filter.  The GSO $\mathbf{S}_\diamond$ in the second sum represents any type of product graph, without loss of generality. This model can be seen as a combination of a G-VAR model for each feature independently, where the graph filter order is zero, with a PG-VAR model, where the graph filter orders of zero are not included. To even further extend the degrees of freedom of~\eqref{eq:first}, we generalize it through the combination of a G-VAR model, which forecasts each feature independently, with a  PG-VAR. We will refer to the combined PG-VAR and G-VAR as the PG-G-VAR model, which is defined as
\begin{align}
\mathbf{x}_{t}=-\left(\sum_{p=1}^{P}\sum_{k=0}^{K-1}\left( \operatorname{diag}(\mathbf{h}_{kp}) \otimes \mathbf{S}^k\right) + \sum_{p=1}^{P} \sum_{k=0}^{K-1} h_{kp} \mathbf{S}_\diamond^{k} \right) \mathbf{x}_{t-p},
\end{align}
where $\mathbf{h}_{kp} = [h^{(1)}_{kp},\hdots,h^{(F)}_{kp}]^\top \in \reals^F$, and $h_{kp}^{(f)}$ is the $k$th order graph filter coefficient associated to the $f$th feature of the $p$th filter.
Compared to a G-VAR per feature, the number of parameters is increased only by a small amount to $PK(F+1)$. This increase comes with the flexibility of modeling each feature separately using the G-VAR and also include information of related features using a type of product graph and the importance of each model is weighted by their graph filter coefficients.

\sn
\textbf{MIMO G-VAR.}
We can further extend the expressiveness of the filter, especially when the feature graph $\bbS_\ccalF$ is not readily available. To this extent, consider an $F$-dimensional graph signal $\bbx_t$, and consider $F$ separate G-VAR filters, all with same orders $P$ and $K$ (without loss of generality), which independently forecast each feature of $\bbx_t$; i.e.:
\begin{equation}
\mathbf{x}_{t}=-\sum_{p=1}^{P} \sum_{k=0}^{K-1}\left( \operatorname{diag}(\mathbf{h}_{kp}) \otimes \mathbf{S}^{k}\right) \mathbf{x}_{t-p}.
\label{eq:G_VAR_MULTI}
\end{equation}
There are a total of $FKP$ coefficients in this model to estimate. To take the influence that features have on each other into account, the diagonal matrix of filter coefficients can be extended into a full matrix with learnable parameters:
\begin{equation}
\mathbf{x}_{t}
=-\sum_{p=1}^{P} \sum_{k=0}^{K-1}\left(\mathbf{H}_{kp} \otimes \mathbf{S}^{k}\right) \mathbf{x}_{t-p},
\label{eq:MIMO-1}
\end{equation}
where the matrices $\mathbf{H}_{kp}\in \mathbb{R}^{F\times F}$ contain the filter coefficients associated to each time-lag $p$ and filter order $k$. To gain more intuition on~\eqref{eq:MIMO-1}, we can rewrite it as:
\begin{equation}
\label{eq:MIMO-2}
\mathbf{X}_t=-\sum_{p=1}^{P} \sum_{k=0}^{K-1} \mathbf{S}^{k} \mathbf{X}_{t-p}\mathbf{H}_{kp},
\end{equation}
where $\mathbf{X}_{t} = [\mathbf{x}_{t}(1),\hdots,\mathbf{x}_{t}(F)]$, i.e., it is the matrix which contains in the $f$th column the one-dimensional graph signal related to feature $f$ at time $t$. From~\eqref{eq:MIMO-2}, it is easy to see that the predicted feature values of $\mathbf{x}_t(f)$ are given by a linear combination of $F$ G-VAR models where each model uses a different feature as input. Another interpretation of the operation in \eqref{eq:MIMO-2} is that the left-multiplication of the data matrix $\bbX_{t-p}$ shifts over the graph $\ccalG$, while the right-multiplication with matrix $\bbH_{kp}$ shifts (averages) over the features of each individual node. This multidimensional graph filtering operation corresponds to a MIMO graph filter \cite{MIMO_GF_2018}, and we refer to this graph-based VAR model as the MIMO G-VAR. This model has a total amount of $PKF^2$ learnable parameters and a computational complexity of $\mathcal{O}(PKF^2\left|\mathcal{E}\right|)$.

\section{Parameter estimation}
\label{sec:param_estimation}
In this section we first elaborate on the least-squares (LS) estimation of the graph filter coefficients, then we propose an estimation method that jointly learns the filter coefficients and the feature graph.

\subsection{Multivariate Least Squares Estimator}
\label{sec:LS}
The first method is the multivariate LS estimator, which is one of the most used methods to estimate VAR coefficients \cite{lütkepohl_2005}. Notice that all the  proposed graph-based VAR models are linear in the filter coefficients; as such, with a proper reshaping of the matrices involved in the filtering operations, they can be estimated in a similar way. For simplicity, we will show the estimation for the G-VAR case. Assume there are $T+P$ multi-dimensional graph signal samples available. Let ${\bf h} \in \mathbb{R}^{KP} $ represent the vector that contains the unknown filter coefficients of the G-VAR. The least-squares estimator to find the filter coefficients that best fit the data is given by the argument that minimizes the sum of squared errors. For the G-VAR this is written as,
\begin{align}
\hat{\mathbf{h}}=  \text{arg}\underset{\mathbf{h}}{\text{ min}}  \sum_{t=1}^{T}\left  \| \mathbf{x}_t + \sum_{p=1}^{P}\mathbf{H}_p(\mathbf{S}) \mathbf{x}_{t-p} \right \|^2_2.
\label{eq:graph-VAR-LS_sum}
\end{align}
where $\mathbf{h}=[h_{01}, \ldots, h_{0P}, \ldots, h_{K1}, \ldots, h_{KP}]^T$. This minimization has a closed-form solution, which for the G-VAR can easily be formulated if the model is rewritten as a matrix-vector product. Using all $T+p$ samples, we define
\begin{align}
\mathbf{X} &=  \begin{bmatrix}
\mathbf{x}_{P-1} &\cdots   &\mathbf{x}_{0} \\ 
\vdots  & \cdots & \vdots \\ 
\mathbf{x}_{T+P-1} &\cdots   & \mathbf{x}_{T}
\end{bmatrix} \nonumber \\
 \mathbf{A}&=\begin{bmatrix}
\left (\mathbf{I}_T\otimes \mathbf{S}^{0} \right) \mathbf{X}  &\cdots   & \left (\mathbf{I}_T\otimes \mathbf{S}^{K-1}  \right) \mathbf{X} 
\end{bmatrix}\\
\mathbf{b}&= [\bbx_{P-1}^\top, \ldots, \bbx_{T+P}^\top]^\top\nonumber
\end{align}
With this notation in place, we can rewrite~\eqref{eq:graph-VAR-LS_sum} as:
\begin{align}
    \hat{\bbh} =\argmin_{\bbh}\|\bbb + \bbA\bbh  \|_2^2,
\end{align}
for which the LS solution is given by:
\begin{equation}
    \hat{\mathbf{h}} = -(\mathbf{A}^T\mathbf{A})^{-1}\mathbf{A}^T\mathbf{b}=-\mathbf{A}^{\dagger}\mathbf{b}.
    \label{eq:direct-LS}
\end{equation}

\subsection{A Joint estimation approach}
\label{sec:Joint_estimation}
When the feature graph is not available from the context of interest, a possible approach is to learn it together with the filter coefficients. To find the feature graph that best suits the product graph-based VAR model, we propose to identify the weights of the GSO by minimizing the forecasting error. As the optimal graph filter coefficients are dependent on the used feature GSO we jointly estimate them, e.g. for the Kronecker PG-VAR this joint problem is defined as:
\begin{align}
\hat{\mathbf{h}}, \hat{\mathbf{S}}_{\mathcal{F}}=\underset{\mathbf{h}, \mathbf{S}_{\mathcal{F}}}{\operatorname{argmin}}
\sum_{t=1}^{T}&\left\| \mathbf{x}_t-\sum_{p=1}^{P}\sum_{k=0}^{K-1} h_{kp} (\mathbf{S}_{\mathcal{F}}\otimes \mathbf{S})^{k} \mathbf{x}_{t-p}\right\|_{\mathrm{2}}^{2}\nonumber\\
\operatorname{s.t.}\quad&\operatorname{supp}(\mathbf{S}_{\mathcal{F}}) \subseteq \operatorname{supp}(\bbS_{\mathcal{F}}^{(0)})
\label{eq:joint_problem}
\end{align}
where $\mathbf{S}_\mathcal{F}^{(0)}$ represents the initial chosen GSO of the feature graph and  $\operatorname{supp}(\cdot)$ is the set of non-zero elements of the argument, i.e., we assume its sparsity pattern is available from the application at hand.  This is a non-convex problem, due to the polynomials of the feature GSO.

In order to tackle the non-convexity of the problem,  we use the method  in~\cite{Joint_20202}, where $\mathbf{h}$ and $\mathbf{S}_\mathcal{F}$ are estimated iteratively using an alternating minimization (AM) approach where the non-convex portion of the problem is solved through the sequential convex programming (SCP) paradigm~\cite{boyd2008sequential}. The pseudo-code that describes our AM approach is given in Algorithm \ref{alg:joint_AM}. In this algorithm, we use the estimate of $\mathbf{S}_\mathcal{F}$ at the $(n-1)$th iteration to find the vector of parameters $\mathbf{h}$, which is obtained by solving
\begin{equation}
\hat{\mathbf{h}}=\underset{\mathbf{h}}{\operatorname{argmin}} \sum_{t=1}^{T}\left\|\mathbf{x}_t-\sum_{p=1}^{P}\sum_{k=0}^{K-1} h_{kp} (\mathbf{S}_{\mathcal{F}}\otimes \mathbf{S})^{k} \mathbf{x}_{t-p}\right\|_{\mathrm{2}}^{2}.
\label{eq:joint_problem_h}
\end{equation}
As shown above, in Section \ref{sec:LS}, this is a linear least-squares problem, which has a closed-form solution as defined in (\ref{eq:direct-LS}). We then use the estimated graph-based VAR parameters at the $n^{th}$ iteration to estimate $\mathbf{S}_\mathcal{F}$. This estimate is found by minimizing the original objective function with respect to the feature GSO:
\begin{align}
\hat{\mathbf{S}}_{\mathcal{F}}=\underset{ \mathbf{S}_{\mathcal{F}}}{\operatorname{argmin}} \sum_{t=1}^{T}&\left\|\mathbf{x}_t-\sum_{p=1}^{P}\sum_{k=0}^{K-1} h_{kp} (\mathbf{S}_{\mathcal{F}}\otimes \mathbf{S})^{k} \mathbf{x}_{t-p}\right\|_{\mathrm{2}}^{2} \nonumber\\ \operatorname{s.t.}\quad&\operatorname{supp}(\mathbf{S}_{\mathcal{F}}) \subseteq \operatorname{supp}(\bbS_{\mathcal{F}}^{(0)})
\label{eq:joint_problem_S}
\end{align}
This problem is still non-convex, but it is a lighter problem than (\ref{eq:joint_problem}). A non-convex method can be used to obtain $\hat{\mathbf{S}}_{\mathcal{F}}$, as the gradient is not hard to find, we use the sequential quadratic programming (SQP) method; see \cite{Joint_20202} for details.

\begin{algorithm}
\caption{Joint GF and feature GSO}
\begin{algorithmic}[1]
\Require $\bbS_{\mathcal{F}}^{(0)}, \epsilon > 0$
\State $n \gets 1$
\While{not converged}
    \State $\mathbf{h}^{(n)} \gets \operatorname{arg}\underset{\mathbf{h}}{\operatorname{min}} f\left( \mathbf{h}, \mathbf{S}_{\mathcal{F}}^{(n-1)}\right) $\Comment{See equation (\ref{eq:joint_problem_h})}
    \State $\mathbf{S}_{\mathcal{F}}^{(n)}\gets \operatorname{arg}\underset{\mathbf{S}_{\mathcal{F}}}{\operatorname{min}} f\left(\mathbf{h}^{(n)}, \mathbf{S}_{\mathcal{F}}\right)$  \Comment{See equation (\ref{eq:joint_problem_S})}
    \State$\text{Check convergence }(\mathbf{h}^{(n)},\mathbf{S}_{\mathcal{F}}^{(n)},\epsilon)$
    \State $n \gets n+1$
\EndWhile
\State \Return $\mathbf{h}^{(n)},\mathbf{S}_{\mathcal{F}}^{(n)} $
\end{algorithmic}
\label{alg:joint_AM}
\end{algorithm}

\section{Numerical Results}
\begin{figure*}
     \centering
     \begin{subfigure}[b]{0.49\textwidth}
         \centering
         \includegraphics[width=0.97\textwidth]{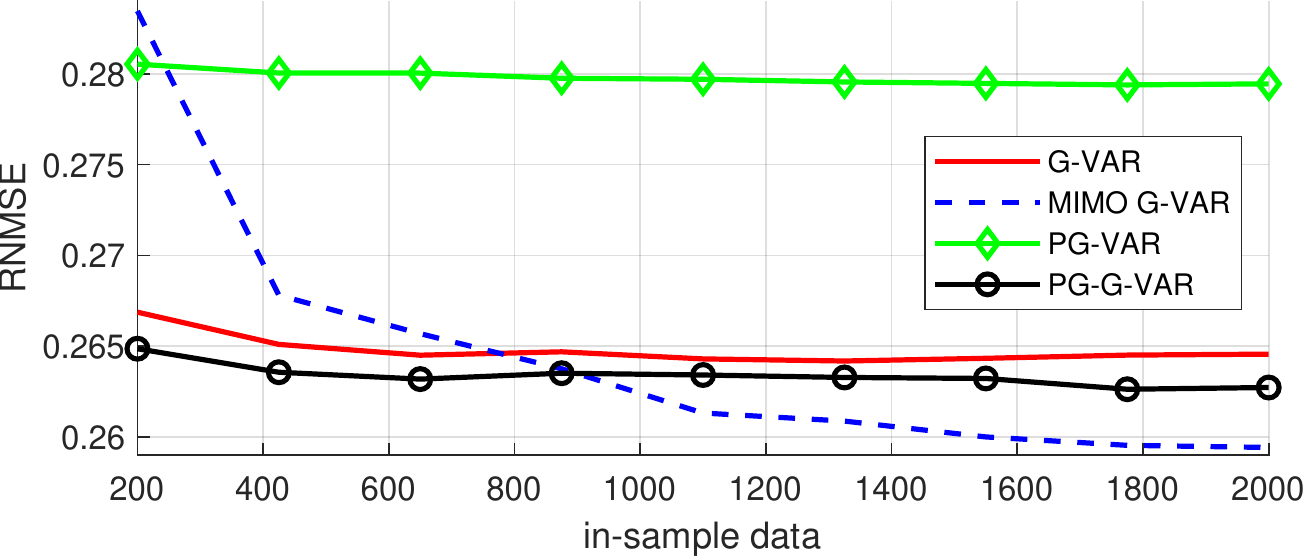}
         \caption{}
         \label{fig:China_non_est}
     \end{subfigure}
     \hfill
     \begin{subfigure}[b]{0.49\textwidth}
         \centering
         \includegraphics[width=0.97\textwidth]{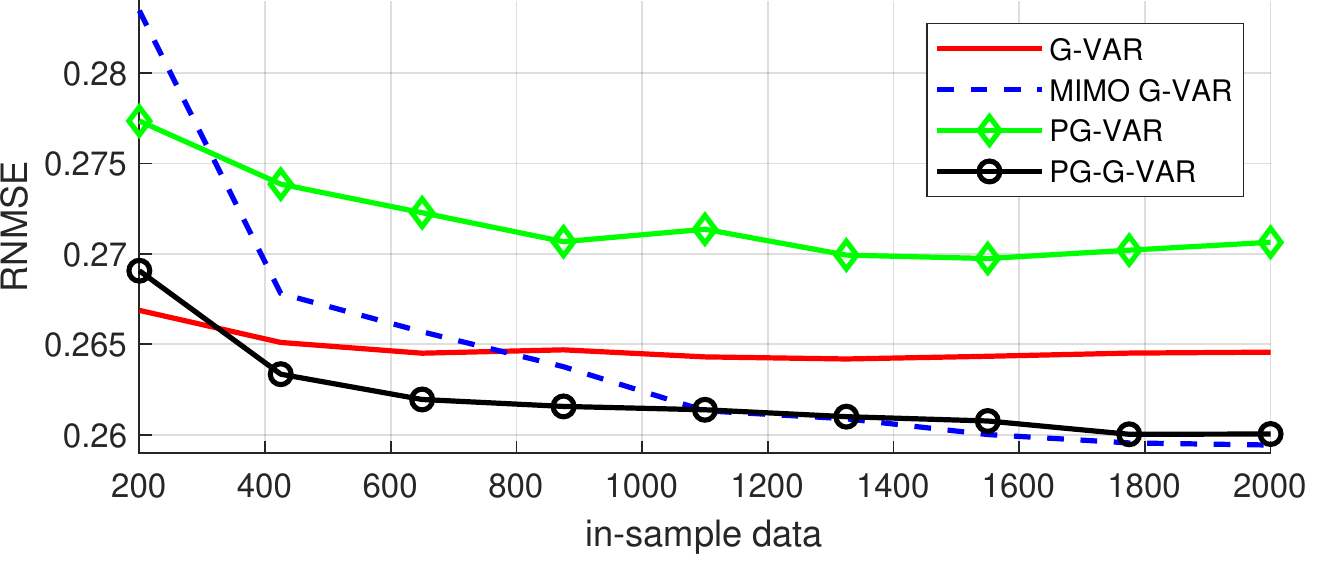}
         \caption{}
         \label{fig:China_estimated}
     \end{subfigure}
        \caption{RNMSE versus the amount of in-sample data for the different graph-based models. Where at (a) the results are shown when the initial feature graph is used and at (b) the joint estimation method is used to find the optimal feature edge weights.}
        \label{fig:Results}
\end{figure*}

To evaluate the models we use a sliding window cross-validation setup \cite{EvalTimeSeries2019}; specifically, the time series are split along the temporal axis into three parts: an in-sample, an out-of-sample, and a left-out part. The in-sample data, which is further split into a training and validation set following a  $70\%/30\%$ scheme~\cite{hastie01statisticallearning}, is used to find the optimal model hyperparameters through a grid search (such as $P$ and $K$) and to estimate the filter coefficients; the out-of-sample data serves as a ``test'' set to measure the prediction accuracy. The left-out part of each iteration consist of the data that is not taken into account. At each iteration the in-sample and out-of-sample parts slide over the data set, creating multiple data sets where the models can be trained and evaluated over. The performance metric considered is the root normalized mean squared error (RNMSE) over the out-of-sample data for all iterations, which is defined as:
\begin{equation}
\text{RNMSE}=\sqrt{\frac{\sum_{t=1}^{\tau}\left\|\tilde{\boldsymbol{\mathbf{x}}}_{t}-\boldsymbol{\mathbf{x}}_{t}\right\|_{2}^{2}}{\sum_{t=1}^{\tau}\left\|\boldsymbol{\mathbf{x}}_{t}\right\|_{2}^{2}}},
\end{equation}
where $\tau$ is the number of signals considered to compute the RNMSE, $\bbx_t$ is the true graph signal, and $\tilde{\bbx}_t$ is the predicted signal.

\subsection{Data set}
We evaluate the performance of the graph-based forecasting models on the Beijing air-quality data set \cite{Zhang2017CautionaryTO,UCIML_data}. There are $F=10$ features considered, 6 types of air pollutants (PM2.5, PM10, SO2, NO2, CO, O3) and 4 weather-related variables (temperature, pressure, dew point, and wind speed), all recorded by $N=12$ air-quality monitoring stations, representing our nodes, in the Beijing area. The data considered is from 20 July 2015 at 7:00 to 5 September 2016 at 13:00, which results in a set of $T=9918$ hourly measurements. The graph regarding the measurement stations is constructed with a 3 nearest neighbors approach, based on the geographical distances and has edge weights  defined by a Gaussian kernel weighting function, similar to~\cite{GVARMA2019}. The feature graph is constructed by connecting each feature to its two most correlated features. We use the normalized Laplacian as GSO for both the station graph and the feature graph.

\subsection{Results}
We evaluate our proposed models , the MIMO G-VAR and the PG-G-VAR, and compare their performance with the PG-VAR and G-VAR. The Cartesian graph product is considered as the product graph type that models the relations between the station graph and the feature graph. The following sets of parameter values are taken into consideration : $P,K \in \{1,\ldots,5 \}$. For all data sets, the range of in-sample data samples considered is from 200 to 2000, and the out-of-sample data consists of 168 hourly measurements, i.e. one week of data. The amount of iterations is 20, and at every iteration, all data is shifted with the number of out-of-sample data points.

First, we evaluate the results with the initial defined feature GSO and do not use the joint estimation method to estimate it. From the results in Fig. \ref{fig:China_non_est}, it can be seen that, as we expected due to its limitations, the PG-VAR model has a decreased performance compared to the G-VAR for each feature. The combined PG-G-VAR model increases the accuracy compared to the G-VAR. The MIMO G-VAR needs more training data since it is the most flexible, but eventually has the best performance. Secondly, in Fig. \ref{fig:China_estimated} the results are shown for the case we do estimate the weights of the feature GSO jointly with the graph filter coefficients. A significant increase in accuracy can be seen for the PG-VAR and PG-G-VAR models. However, although the PG-VAR shows an improvement it still performs less than the other methods. The joint estimation method results for the PG-G-VAR model show that it outperforms all other methods for smaller amounts of in-sample data. Except for the 200 in-sample data size where it has a similar performance as the G-VAR, and for larger amounts of in-sample data where it has now a similar performance as the MIMO G-VAR model. This illustrates the importance of estimating the edge weights of the feature graph when a product graph is applied. Further, it showcases the advantage offered by using a priori knowledge of the feature graph, which makes it work well with lower amounts of in-sample data.

\section{Conclusion}
In this paper, we proposed two new models to forecast multi-dimensional graph signals, the PG-G-VAR and MIMO G-VAR models.
Further, we applied a joint estimation approach to estimate the graph filter coefficients together with the weights of the feature graph. The results indicate that our limitation assumptions regarding the PG-VAR are correct, as it shows a reduced prediction performance compared to the G-VAR model per feature. On the other hand, the results show a clear improved prediction accuracy for the models introduced in this paper, especially for more training data. 

Further work is needed to enhance the proposed PG-G-VAR model's estimation cost. Especially in the direction of finding optimal weights of the feature graph, as the numerical experiments showed that estimating $\mathbf{S}_\mathcal{F}$ leads to increased forecasting performance, but with the cost that a non-convex problem has to be solved. A possible solution for this would be to apply a constrained edge-variant graph filter ~\cite{disertation}, to the feature graph. In this filter, each element in the GSO is weighted individually by a graph filter coefficient, before it is used in the filtering operation. This results in a linear model with respect to these graph filter coefficients.

\newpage

\bibliographystyle{IEEEbib}
\bibliography{refs}

\end{document}